\documentclass[aip,jcp,amsmath,amssymb,
reprint]{revtex4-1}
\usepackage{graphicx}
\usepackage{dcolumn}
\usepackage{bm}
\usepackage{array,mathtools,amssymb,booktabs}
\usepackage{lmodern} 
\usepackage{upgreek}
\usepackage[hidelinks]{hyperref}

\begin{document}
\title{Phase Separation and Aggregation in Multiblock Chains}

\author{Athanassios Z. Panagiotopoulos}
\email[email:]{azp@princeton.edu}
\affiliation{Department of Chemical and Biological Engineering, Princeton University, Princeton, NJ 08544, U.S.A.}
\date{\today} 

\begin{abstract}
This article focuses on phase and aggregation behavior for linear chains composed of blocks of hydrophilic and hydrophobic segments. Phase and conformational transitions of patterned chains are relevant for understanding liquid-liquid separation of biomolecular condensates, which play a prominent role in cellular biophysics, but also for surfactant and polymer applications.  Previous studies of simple models for multiblock chains have shown that, depending on the sequence pattern and chain length, such systems can fall into one of two categories: displaying either phase separation or aggregation into finite-size clusters. The key new result of the present study is that both formation of finite-size aggregates and phase separation can be observed for certain chain architectures at appropriate conditions of temperature and concentration. For such systems, a bulk dense liquid condenses from a dilute phase that already contains multi-chain finite-size aggregates. The computational approach involves several distinct steps using histogram-reweighting grand canonical Monte Carlo simulations, which are described here in some level of detail. 
\end{abstract}

\maketitle

\section{Introduction}

Liquid-liquid phase separation in biological systems is  considered a key mechanism for subcellular organization, with important implications to the functioning of living cells.\cite{Bra09,Ban17,Shi17} More broadly, formation of two liquid phases by polymers in both aqueous\cite{Hat01} and non-aqueous\cite{Dob80} solvents plays an important role for separation and purification technologies and provides a challenging domain for thermodynamic and statistical mechanical modeling methods. Similarly, formation of finite-size aggregates (micelles) from both low-molecular weight surfactants\cite{Ros12} and macromolecular components\cite{Rie03} is important for detergency, catalysis, and nanotechnology applications. 

While  sophisticated models have been proposed for describing biomolecular condensation \cite{Dig18,Per21}, significant insights have also been obtained from simple models containing a small number of residue types. Such models have also proven useful in modeling non-biological two-phase systems and in understanding micellization in surfactant and block copolymer solutions. For example, a linear chain model with just hydrophobic and hydrophilic beads has been deployed to illustrate the sensitivity of phase behavior to sequence and to generate many complex condensate morphologies\cite{Sta20, Sta21}, using large-scale molecular dynamics simulations. A significant limitation of these simulations is that the precise character of the observed transitions (e.g., first-order versus continuous) cannot be determined. In order to overcome these issues and to investigate the interplay between aggregation and phase separation, it is important to have access to accurate estimates of the free energy of a system as a function of thermodynamic state variables over a broad range of concentrations and temperatures. For this purpose, a particularly useful approach involves utilizing a lattice chain model originally proposed by Larson et al.\cite{Lar85} The model was designed to represent chains with  hydrophilic (head, ``H'') and hydrophobic (tail, ``T'') segments and has provided insights on micellization and phase transitions of diblock and triblock surfactant chains,\cite{Flo99,Pan02} multiblock chains consisting of identical repeat units,\cite{Gin08} and has been used to model liquid-liquid phase separation in disordered proteins.\cite{Ran21}

Prior work on aggregation and phase behavior of lattice chains with hydrophobic and hydrophilic beads has produced some general trends on the effects of sequence and chain length on the observed properties. The ``all-T'' version of the model is identical to the well-known lattice homopolymer system underlying Flory-Huggins theory.\cite{Flo53} It has a first-order transition between dense and dilute fluid phases, for which the critical temperature increases and critical volume fraction decreases as chain length is increased. Short diblock chains such as H$_2$T$_4$ have been shown to form finite-size aggregates in a continuous transition, while longer ones, such as H$_2$T$_8$ have a first-order dilute-dense fluid phase transition.\cite{Pan02} For triblock chains, hydrophilic groups at the end favor micellization, whereas hydrophobic end groups favor phase separation; once again, longer sequences of T blocks lead to phase separation. For example the sequence T$_2$H$_8$T$_2$ micellizes, while T$_4$H$_8$T$_4$ phase separates. Multiblock chains follow the same pattern of chain length favoring phase separation: e.g., (H$_3$T$_3)_3$ forms finite-size aggregates, while (H$_3$T$_3)_4$ phase separates.\cite{Gin08} In the most recent study of this model,\cite{Ran21} a large number of sequences with lengths between 20 and 100 beads were examined and classified as phase separating or aggregating. In that study, sequences with long blocks of T beads were determined to phase separate while sequences with shorter blocks phase separate. As chain length increased, a large majority of long sequences was determined to phase separate, a fact of possible biological relevance for disordered proteins. 

Interestingly, all prior studies mentioned in the previous paragraph were not able to observe both formation of finite-size aggregates and phase separation for any specific chain architecture. This poses the interesting question of the interconnection between the different types of transition and how one type of behavior changes to another for systems ``on the cusp'' between aggregation and phase separation. In the present study, we aim to address this question by revisiting some of these previously studied systems reported to be at the boundary between the two types of behavior, using new simulations combined with careful analysis of the results.

Specifically, we use grand canonical Monte Carlo simulations combined with histogram reweighting for determination of phase coexistence curves, critical points, and aggregation boundaries. This approach is now quite mature, having served the community well for about twenty-five years. However, there are several subtle points in implementation of the methodology for systems that may have multiple types of transitions. For this reason, this article contains a pedagogical component with a relatively detailed description of the computational approach, in addition to presentation of the new findings. The codes as well as example input and output files are available online, as detailed in the Data Availability Statement. 

\section{Model and Methods}

\subsection{Model}

\begin{figure}
    \centering
    \includegraphics[width=0.8 in]{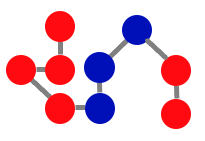}
    \caption{A configuration for the triblock T$_4$H$_3$T$_2$ chain with energy $E=-2$. For simplicity, a 2-dimensional snapshot is shown, even though all calculations were performed in 3-dimensional space.}
    \label{fig:model}
\end{figure}

The model used in this study is shown schematically in Fig. \ref{fig:model}. It consists of linear chains on a simple cubic lattice, entailing two types of beads, namely hydrophobic T (red color) and hydrophilic H (blue color). Excluded volume interactions are active among beads, so that only a single one can occupy any given lattice site. Chain connectivity is along the main directions of the lattice, with bonds between bead centers of length 1 in lattice units, but also diagonal  in-plane bonds of length $\sqrt2$, and diagonal off-plane bonds of length $\sqrt3$. In other words, bond directions can be [0 0 1], [0 1 1], [1 1 1] and their allowable rotations and reflections, resulting in 26 total possible connectivity vectors. These same vectors define nearest-neighbor attractive T--T interactions for non-bonded beads, of magnitude $-1$, in reduced units that also set the temperature scale. All other interactions, namely between H--T and H--H beads, as well as interactions involving the monomeric solvent beads occupying space not containing chains, are set to zero. In some early work on this model,\cite{Flo99,Pan02,Gin08} the interaction energy scale was set to $-2$ to be consistent with the original publications.\cite{Lar85,Lar88} This would necessitate a trivial scaling by a factor of 2 of temperatures, energies, and chemical potentials in the present article relative to the results reported in Ref. \citenum{Ran21}. The model describes liquid-liquid separation between  solvent-rich and solvent-lean phases in a two-component system of solvent and chains on a fully occupied lattice. A completely equivalent description of the model would be that it describes vapor-liquid equilibria of the one-component system of chains in a partly occupied lattice. 

\subsection{Grand Canonical Monte Carlo}

Monte Carlo sampling in the grand canonical ensemble was used to obtain all aspects of thermodynamic behavior for the lattice chain model described in the previous subsection.  For a one-component system, the control variables of this ensemble are the chemical potential $\mu$, volume $V$ and temperature $T$. For the two-component ``filled lattice'' equivalent description of the model, the first control variable is the chemical potential difference between chains and a number of monomers equal to the chain length, but here we adopt the simpler one-component notation. Cubic boxes of edge length $L$ were simulated, so that the total number of sites was $V=L^3$. At least two different box sizes were used for each system studied, in order to quantify finite-size effects on the results.

The elementary Monte Carlo moves used in the simulations were chain insertion or removal attempts (typically 60 \% of total moves) partial chain regrowth (39.8 \% of total moves), and cluster moves (0.2 \% of total moves).  The regrowth moves involve cutting a chain at a randomly selected internal bond and regrowing it from that point to the end of the chain. To facilitate insertion / removal and partial chain regrowth moves, an athermal Rosenbluth algorithm\cite{Ros55} was implemented, as detailed in Ref. \citenum{Pan98}. Cluster moves involved identifying a set of interacting chains and displacing the cluster by a single lattice unit in a randomly chosen spatial direction.\cite{Flo99} These moves help with equilibration at low densities when aggregates are present, but become inefficient and computationally expensive at higher densities for which system-spanning large clusters frequently exist. Typical runs consisted of $2 \times 10^9$ Monte Carlo moves and required a few hours of wall-clock time on a single core of a 3.6 GHz Intel Core i9 processor. The source codes, example input and output files, and a list of the runs performed for each system are available online, as explained in the  Data Availability Statement at the end of this article. Statistical uncertainties were obtained from the standard deviation of results from four independent runs at identical conditions. 

Histogram reweighting was used to allow for full utilization of information from the grand canonical simulations, thus minimizing the number of separate simulation runs required. This method relies on the Ferrenberg-Swendsen algorithm\cite{Fer89} for combining information from separate runs that have a reasonable degree of overlap in the number of particles $N$ and energy $E$ covered in each. The combined histograms are then used to obtain distributions of particle numbers and energy, $P(N,E)$, as a continuous function of $\mu$ and $T$. This capability is key to the determination of phase diagrams and critical points, as explained in the subsection that follows. Histogram reweighting methods used here are described in more detail in a review article.\cite{Pan00}  Computational requirements for histogram reweighting increase as $N^{3/2}$, so it is beneficial to operate at small system sizes, especially when exploring the behavior of a new chain sequence for which the type of transition and location of a possible critical point are not known. This scaling results from the fact that relative fluctuations in density, which are responsible for histogram overlap, are proportional to $N^{-0.5}$, so more runs are needed when bigger boxes are used. Also, sampling needs to be at least proportional to $N$, to allow for a similar number of moves per particle in the simulations. 

\subsection{Determination of Phase Diagrams and Critical Points}
\begin{figure}
    \centering
    \includegraphics[width=3.2 in]{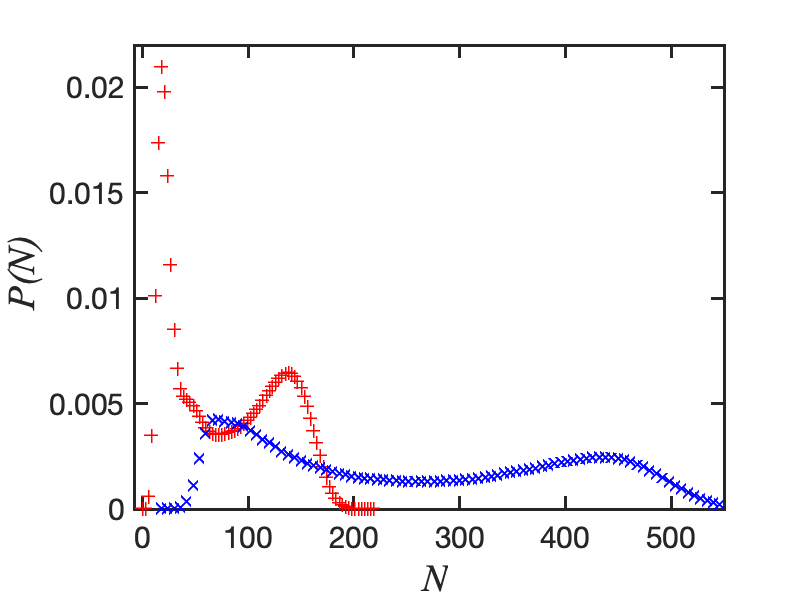}
    \includegraphics[width=3.1 in]{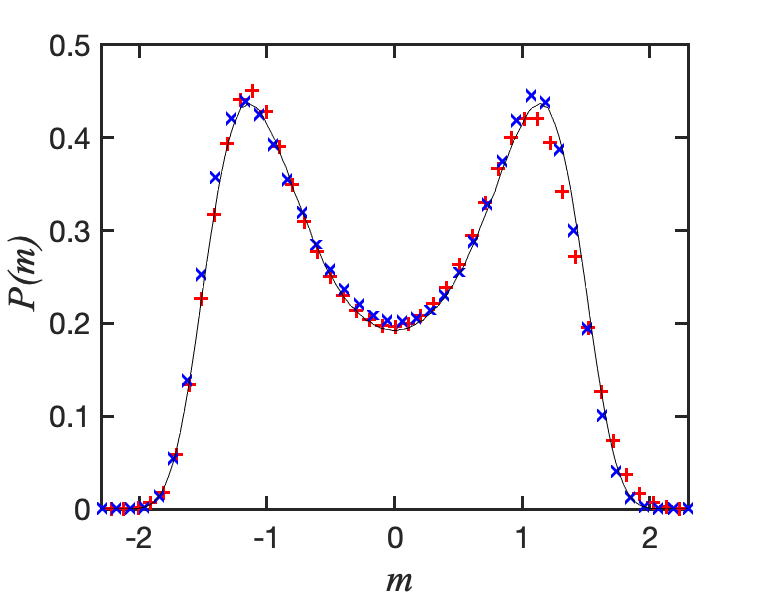}
    \caption{(top) Probability functions $P(N)$ for T$_4$H$_8$T$_4$ at conditions corresponding to the critical point. Red $+$'s are for $L=20$ and blue $\times$'s for $L=30$.  (bottom) Transformed probability functions $P(m)$ for the same system and conditions. The thin black line is the universal three-dimensional Ising order parameter distribution from Ref. \citenum{Tsy00}.}
    \label{fig:P_N}
\end{figure}

For a macroscopic system at equilibrium, one expects an abrupt jump in the density (and thus also the number $N$ of particles in the simulation box)  at subcritical temperatures, occurring at precise value of the chemical potential $\mu$. However, in grand canonical simulations of finite systems, these transitions are subject to significant hysteresis because of the free energy barriers for formation of a new phase. Hysteresis disappears at sufficiently high temperatures where  barriers are small and runs can sample a range of densities covering both coexisting phases. For example, this is the case for the runs underlying the probability functions $P(N)$ shown in Fig.~\ref{fig:P_N} (top). Such runs can be combined through histogram reweighting with subcritical runs of the separate phases to construct coexistence curves. This must be done in a self-consistent manner: the temperature - chemical potential curve at coexistence for a given system must be consistent with the underlying runs used to construct it. In other words, a dense phase simulation must have been  performed at a chemical potential close to, but a little greater than, the chemical potential at coexistence at that temperature. A dilute-phase simulation must have been  performed at a chemical potential a little lower than the chemical potential at coexistence at the temperature of interest. In practice, only a small number of new runs are required to achieve self-consistency. 

Coexistence is  determined by the condition of area equality for the low-$N$ and high-$N$ regions of the $P(N)$ distributions. This construction is meaningful only at lower temperatures than the ones shown in Fig.~\ref{fig:P_N} (top): in the figure, the two peaks potentially corresponding to coexisting phases have a great deal of overlap. At lower temperatures, however, there is a broad range of intermediate $N$ values for which the probability $P(N)$ is low, resulting in clear separation between high-density and low-density peaks. It should be emphasized here that these deeply subcritical distributions cannot be sampled directly, as a single grand canonical run can only sample one phase, because of the barriers to nucleation of the other phase. Only the combination of these runs with near-critical ones that sample both sides makes it possible to determine coexistence (binodal) curves.

To determine critical points and to provide unequivocal confirmation that an apparent phase transition is truly a first-order one between macroscopic phases, the mixed-field finite-size scaling method of Bruce and Wilding \cite{Bru92,Wil92} is used. The method transforms the two-dimensional histograms $P(N,E)$ into one-dimensional probability distributions $P(m)$, where the order parameter for the transition is now $m=N-sE$. The critical chemical potential $\mu_c$, critical temperature $T_c$, and ``field mixing'' parameter $s$ are determined for each chain architecture and system size $L$ from the combined histograms obtained at appropriate state conditions, usually near the critical point. A typical two-dimensional frequency distribution $P(N,E)$ for a system at coexistence is available as Fig. 5 in Ref. \citenum{Pan00} -- the full two-dimensional distribution needs to be used for obtaining the critical parameters. Probability distributions of particle numbers, $P(N)$ for two different system sizes for chains of type T$_4$H$_8$T$_4$ are shown in Fig.~\ref{fig:P_N} (top). As can be seen in the figure, the distributions are asymmetric, with the one for larger system size extending over a broader range of particle numbers.  The transformed distributions $P(m)$ are shown in Fig.~\ref{fig:P_N}(bottom) and compared to the universal Ising three-dimensional distribution. While the fit is not perfect, especially for the smaller system size, it is clear that for this particular system the order parameter probability distributions match the universal curve quite well.

\subsection{Determination of Critical Micellar Concentrations}
\begin{figure}
    \centering
    \includegraphics[width=3.3 in]{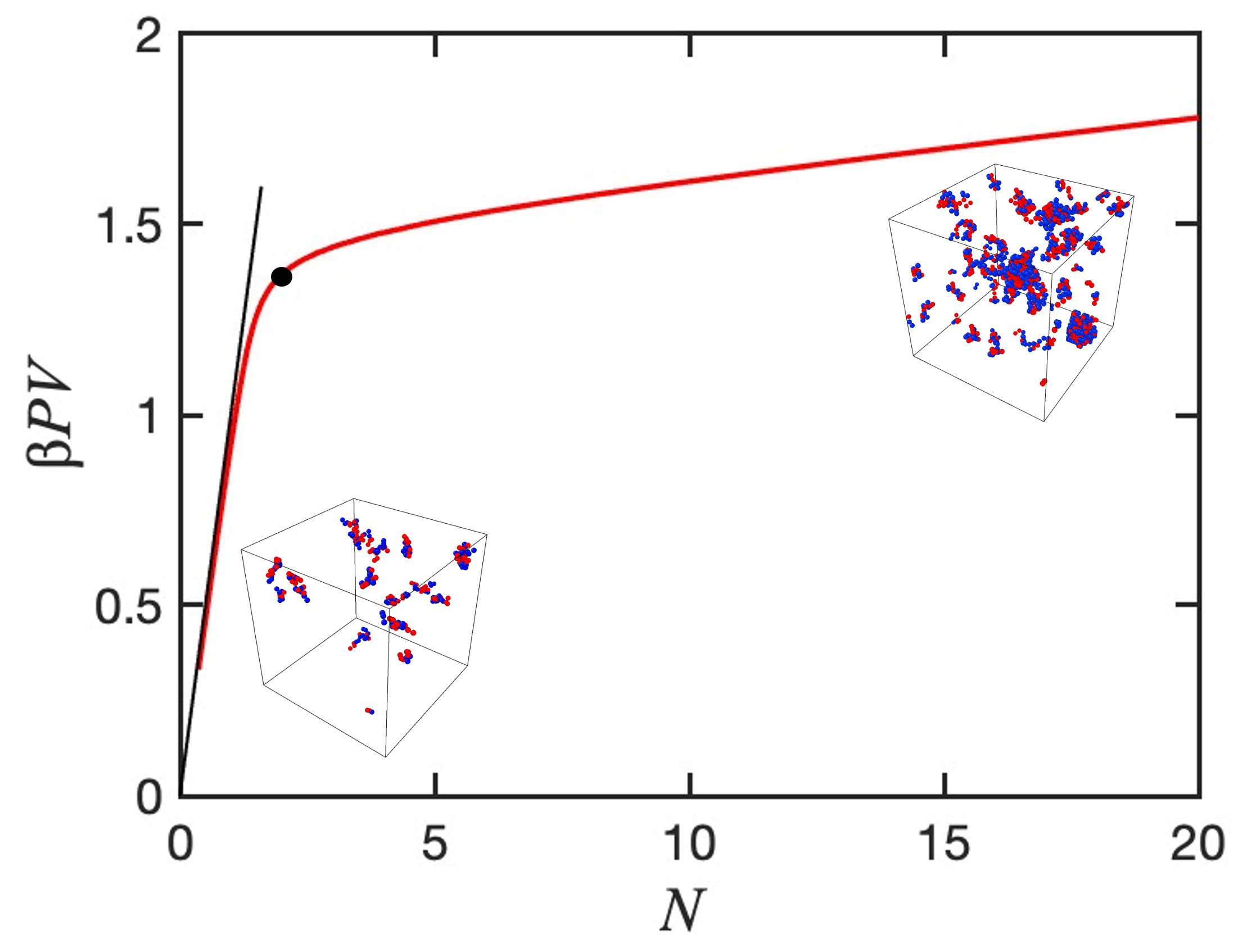}
    \caption{(main plot) The combination $\upbeta PV$ as a function of $N$ for T$_4$H$_8$T$_4$ at $T=3.2$ for $L=20$. A black dot marks the position of maximum second derivative and thus the assigned cmc. The thin black line is drawn with unit slope. (insets) Snapshots of configurations in a box with $L=50$ at conditions below and above the cmc.}
    \label{fig:cmc}
\end{figure}
Following earlier work,\cite{Flo99} we use  equation of state data obtained through histogram reweighting to pinpoint the location of the critical micellar concentration (cmc), which marks the appearance of multi-chain aggregates in systems of suitable size. The grand canonical partition function $\Xi$ is obtained directly from histogram reweighting for the range of particle numbers $N$ covered by the combined histograms, within an unknown additive constant. Given that $\text{ln}{\Xi}=\upbeta PV$, where $\upbeta=1/k_\text{B}T$, a plot of this quantity versus $N$ has a linear portion of unit slope at sufficiently low densities at which systems are close to an ideal gas and chain monomers dominate the dilute phase, as shown in Fig.~\ref{fig:cmc}. At higher densities and at sufficiently low temperatures, chain interactions start being active and the slope of the $\upbeta PV$ versus $N$ curve becomes markedly lower, signifying that the independent kinetic entities in the system are no longer monomers, but rather aggregates of multiple chains. The point at which this curve has a maximum second derivative (rate of change of the slope), is designated as the cmc. Interestingly, the location of this transition obtained from the maximum in the second derivative is insensitive with respect to simulation box size. In relatively small boxes, such as the one with $L=20$ for which Fig.~\ref{fig:cmc} is drawn, there aren't enough chains in the system at the cmc to form aggregates of the size that would appear in larger boxes. However, the insets to Fig.~\ref{fig:cmc} show representative snapshots of this system in boxes with $L=50$ and clearly show the presence of multichain aggregates at concentrations above the cmc. 

It should be pointed out here that the sharp change in slope of the $\upbeta PV$ versus $N$ curves is not unique to systems forming aggregates -- a first-order phase transition is also associated with a sharp decrease in slope, with the final value of the second segment depending on system size. Thus, a confirmation of the character of the transition (micellization into finite-size aggregates, or formation of a bulk condensed phase), needs to rely on cluster size distribution analyses at conditions beyond the computed cmc as well as on the determination of system-size effects on the equation of state.

\section{Results and Discussion}
As suggested earlier, we focus here on chain architectures that were identified in previous work\cite{Flo99,Gin08} to be at the boundary of phase separation (formation of macroscopic fluid phases) and micellization (formation of finite-size aggregates). The key questions we would like to address are the interplay between the two transition types and the possible identification of chain architectures for which both transitions are present at different thermodynamic conditions. 

\subsection{Systems exhibiting only Phase Separation}
\begin{figure}
    \centering
    \includegraphics[width=3.2 in]{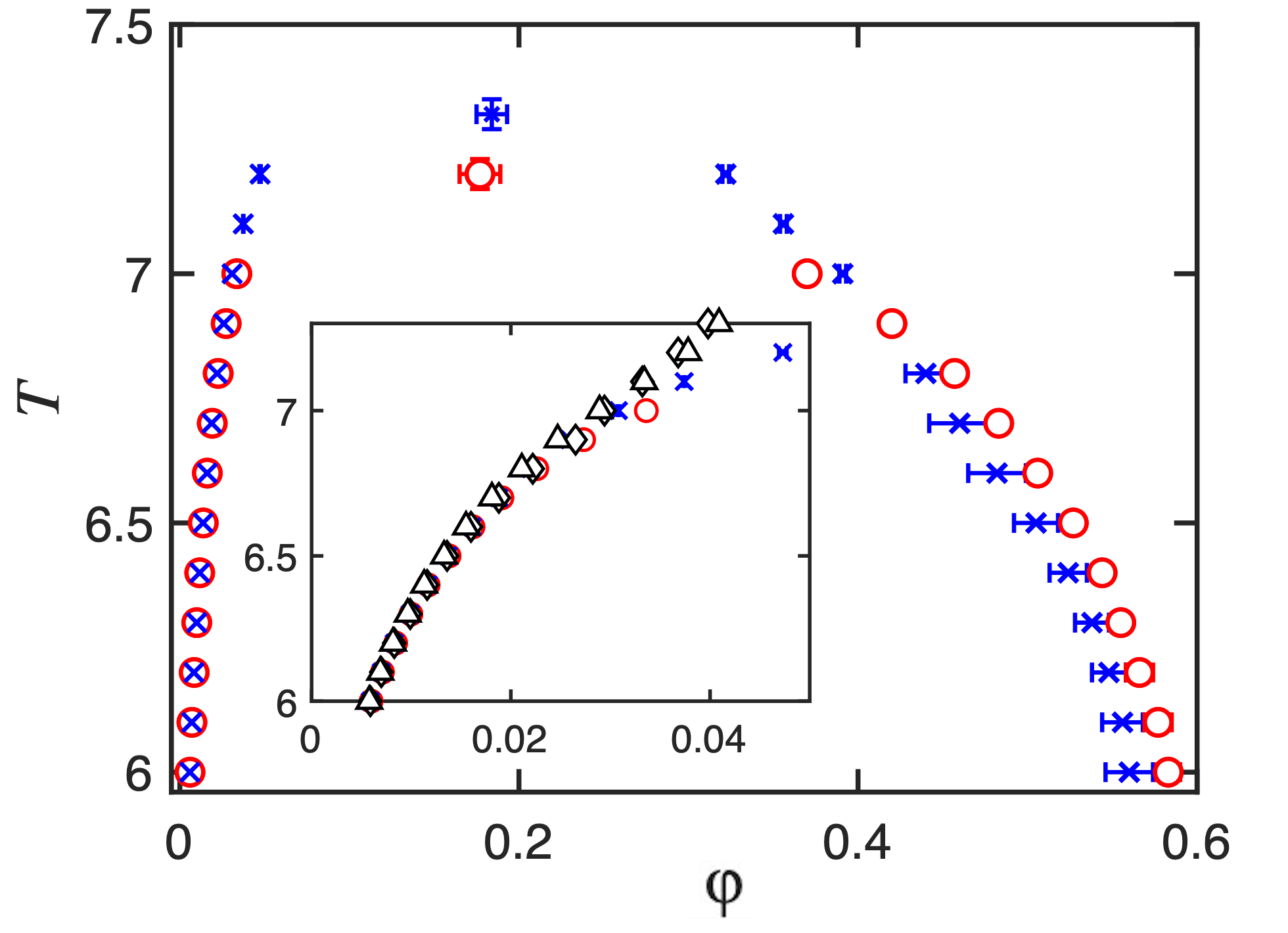}
    \caption{Phase behavior for H$_2$T$_8$. The main plot shows volume fractions at coexistence and critical points for $L=30$ (red circles) and $L=20$ (blue $\times$'s). The inset magnifies the low-density region and also shows apparent cmc's for $L=30$ (triangles) and $L=20$ (diamonds). Statistical uncertainties in this and subsequent figures are visible only when greater than symbol size.}
    \label{fig:H2T8}
\end{figure}

Diblock chains with long sections of solvophobic T beads and short sections of solvophilic H beads undergo phase separation, while longer sections of H beads favor micellization into finite-size clusters protected from further aggregation by the H blocks that cover the exterior of the clusters. The system  H$_2$T$_8$ was previously identified\cite{Flo99} as being on the boundary between phase separation and micellization, but no coexistence data were reported in the prior study.  The phase behavior for this system from the present work is shown in Fig.~\ref{fig:H2T8} for two different system sizes, in terms of temperature $T$ versus volume fraction $\upvarphi=nN/V$, where $n$ is the number of beads per chain ($n=10$ for H$_2$T$_8$). The main plot shows binodal curves and critical points determined as detailed in Section II.C. These binodal curves and critical points are seen to be insensitive to system size, which confirms that the system has a normal bulk liquid phase separated from the dilute phase by a first-order phase transition. 

The inset to Fig.~\ref{fig:H2T8} shows the low-density coexistence region and apparent cmc's, that is, loci of maximum second derivative in the $\upbeta PV$ versus $N$ curves. The apparent cmc's are insensitive to box length $L$ and occur at densities very near the corresponding dilute phase boundary. Additional simulations in large boxes ($L=60$) in the small ``gap'' region between the apparent cmc's and the saturation (binodal) curve showed no micellar aggregates. Instead, the probability distributions decay monotonically from monomers to dimers, trimers, etc, as expected in any system in which attractive interactions are present. For brevity, these distributions are not shown here, but are available online as indicated in the Data Availability Statement. Thus, H$_2$T$_8$ chains show no sign of formation of stable finite-size micellar aggregates, just as is the case for pure T (homopolymer) chains. The presence of the 2-bead hydrophilic segment results in significant reductions, of around 40\%, for both the critical temperature and critical volume fraction relative to the T$_8$ homopolymer values reported in Ref. \citenum{Pan98}.  A reduction of the critical volume fraction for sequences near the boundary between phase separation and aggregation was also observed in Ref. \citenum{Ran21} for general chain sequences and for a variety of chain lengths.

\begin{figure}
    \centering
    \includegraphics[width=3.2 in]{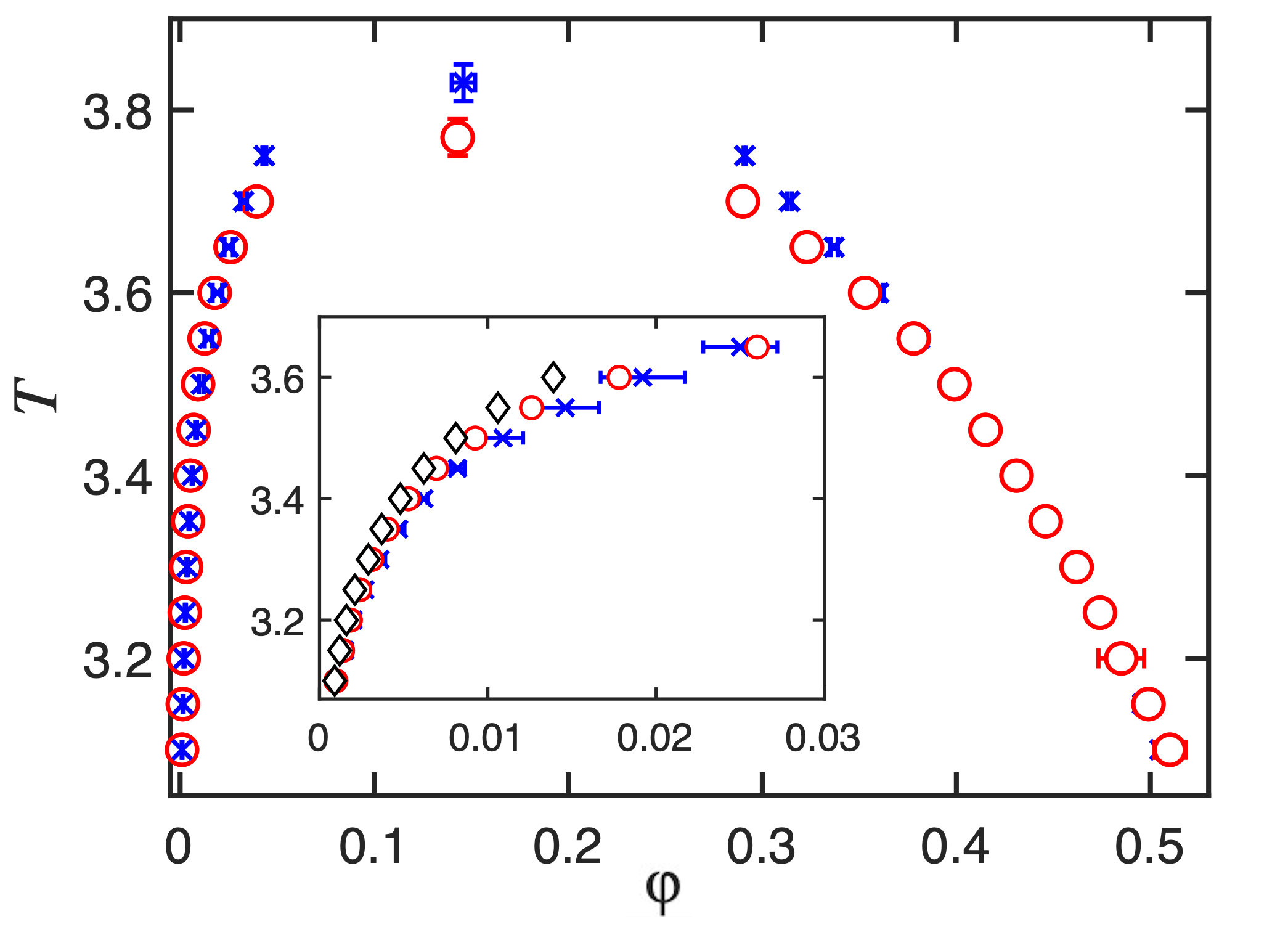}
    \caption{Phase behavior for (H$_3$T$_3$)$_4$. Notation is the same as for Fig.~\ref{fig:H2T8}.}
    \label{fig:H3T3_4}
\end{figure}

A similar ``exclusively phase separation'' behavior is obtained for a multiblock chain, (H$_3$T$_3$)$_4$, previously identified as being on the boundary between phase separation and aggregation.\cite{Gin08} The phase diagram for this system from the present work is shown in Fig.~\ref{fig:H3T3_4}. As was the case for the previous H$_2$T$_8$ system, simulations in large boxes ($L=80$) in the ``gap'' region between the apparent cmc's and the saturation (binodal) curve showed no micellar aggregates. The cluster probability distribution function decayed monotonically from monomers to oligomers, with no separate micellar peak, but in this case had a long tail, with some larger aggregates of dozens of chains at temperatures near the critical point for the first-order transition. The critical temperature for the (H$_3$T$_3$)$_4$ system is about half that of H$_2$T$_8$, despite the greater chain length and the presence of 50\% more attractive T centers on each chain for the former system.

Even though a limited number of chain architectures were examined here to provide examples of systems that display only phase separation and show no signs of aggregation in the dilute phase prior to saturation, it is anticipated that the vast majority of systems identified in prior studies as phase separating fall in this category. This includes chain architectures without extended hydrophilic blocks, as well as long chains of almost any sequence, except the most blocky ones. Formation of finite-size aggregates is the exception, rather than rule, occurring only when there is sufficient ``protection'' of solvophobic segments from further aggregation by sufficiently numerous solvophilic groups on the exterior of any multichain aggregates that form in the dilute phase. 

\subsection{Systems with both Phase Separation and Aggregation}
As stated earlier, prior studies had failed to identify chain architectures displaying both phase separation and aggregation. The reason for this failure turns out to be the fact that chain architectures and lengths for which this dual behavior is accessible from simulations are quite limited. This ``dual'' behavior only occurs over a relatively narrow range of thermodynamic conditions. In the present study, we were able to identify such systems by examining specific chain architectures in large boxes at chemical potentials corresponding to conditions between the cmc and the binodal curve. Specifically, such behavior was identified for two triblock chain architectures and a multiblock one, as follows.  

\begin{figure}
    \centering
    \includegraphics[width=3.5 in]{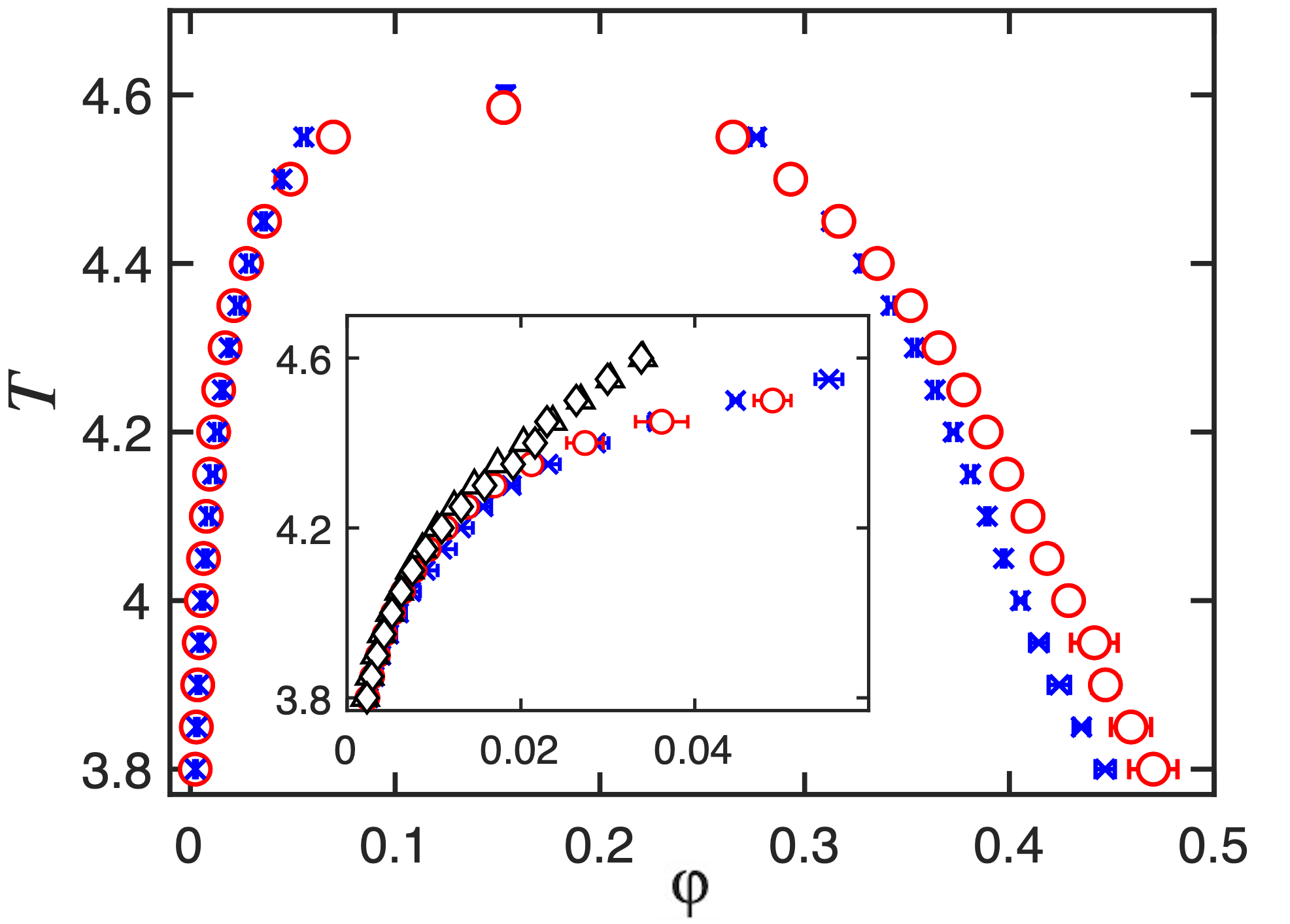}
    \caption{Phase behavior for T$_4$H$_8$T$_4$. Notation is the same as for Fig.~\ref{fig:H2T8}.}
    \label{fig:T4H8T4}
\end{figure}

\begin{figure}
    \centering
    \includegraphics[width=3.5 in]{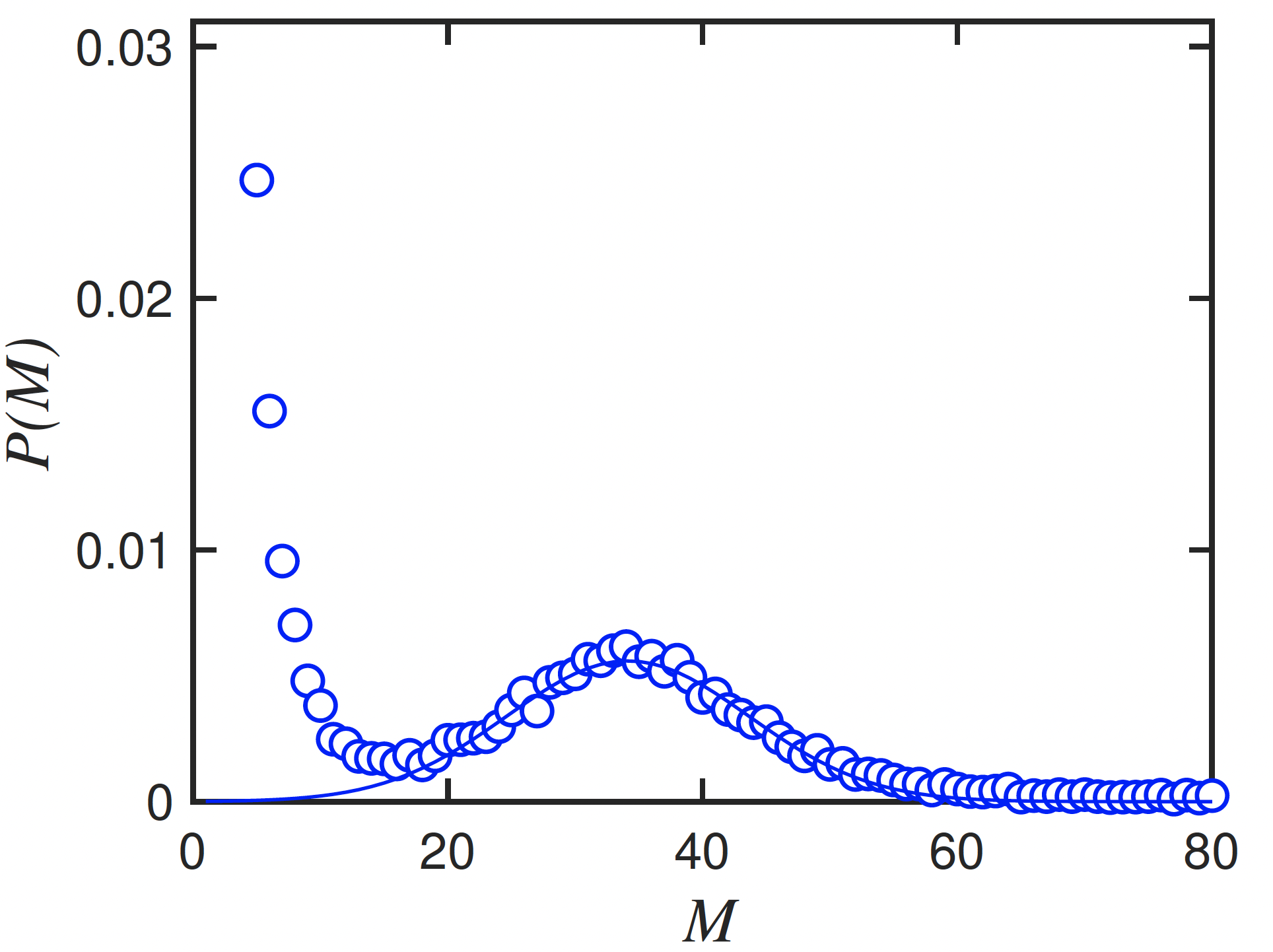}
    \caption{Aggregation number probability function $P(M)$ for T$_4$H$_8$T$_4$, in a box of $L=60$, at $T=4.4$, $<\upvarphi>=0.025$. The line is a Gaussian fit to the data points between $M=20$ and $M=70$.}
    \label{fig:Pm}
\end{figure}

The first system determined to have dual behavior is the triblock chain architecture T$_4$H$_8$T$_4$, which had been previously identified as phase separating, but being close to the boundary for aggregation behavior.\cite{Pan02} The phase behavior computed here is shown in Fig.~ \ref{fig:T4H8T4}. As seen in the inset to the figure, there appears to be a gap between the concentration at which the cmc is assigned from the maximum in the second derivative of the  $\upbeta PV$ versus $N$ curves and the coexistence (binodal) curve on the dilute phase side. As also seen earlier, the cmc location is insensitive to system size. Relatively small boxes of $L=20$ or $L=30$ are optimal for cmc and coexistence curve calculations, as they afford good overlap between runs at different conditions through histogram reweighting. However, in these smaller boxes, at the coexistence density, there are relatively few chains present in the simulation box at any given time, not enough to form micellar aggregates. For this reason, it is necessary to examine large boxes of $L=50$ or more, to find out if there are aggregates in the vapor phase. This needs to be done at chemical potentials between that corresponding to the cmc and phase coexistence. Probability distributions $P(M)$, where $M$ is the size of aggregates, are determined through a cluster algorithm as detailed in Ref. \citenum{Flo99}. This function is shown in Fig.~ \ref{fig:Pm} for a box of $L=60$ at concentrations within the dilute phase, above the cmc but fairly close to coexistence. A well-separated micellar peak can be seen in the data. At these conditions, about 15 \% of the chains in the dilute phase participate in micellar aggregates, which have an average size $<M>=34$.

\begin{figure}
    \centering
    \includegraphics[width=3.5 in]{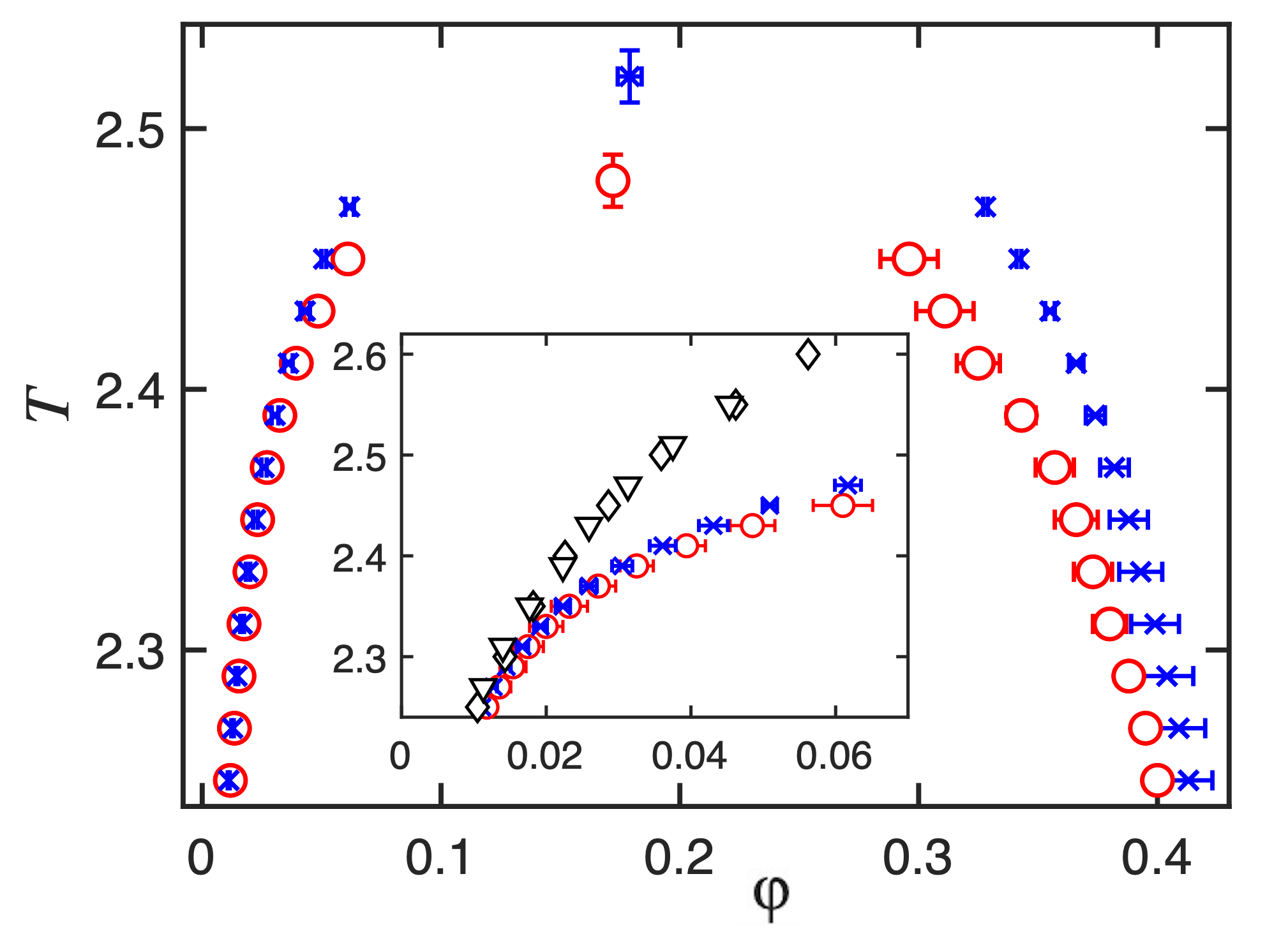}
    \caption{Phase behavior for T$_2$H$_8$T$_2$. Notation is the same as for Fig.~\ref{fig:H2T8}.}
    \label{fig:T2H8T2}
\end{figure}

The second system found to display both aggregation and phase separation is also a triblock chain, namely T$_2$H$_8$T$_2$. This system was identified in\cite{Pan02} to be aggregating. Physically, we expect that the higher proportion of H blocks will result in a more protected hydrophobic core of the aggregates relative to T$_4$H$_8$T$_4$, promoting stable aggregates. The phase envelope and cmc obtained in the present work are shown in  Fig.~\ref{fig:T2H8T2} -- these cmc's are in good agreement with results from.\cite{Pan02} As for the previous triblock system, we find both aggregation and phase separation, but now with a significantly broader gap between the cmc and the phase boundary. There is also a more pronounced system-size dependence of the phase envelope. The critical temperature is about half that of the previous case, as expected from the smaller number of T beads in the attractive end segments. Simulations of the dilute phase at near-coexistence conditions in large boxes  show micellar aggregates for this system, even at temperatures above the critical point for the first-order transition. For example, at the supercritical temperature of $T=2.60$, a run with $<\upvarphi>=0.09$, which is 50\% higher than the cmc, had a well-separated micellar peak with $<M>= 30$ and about 1/3 of the chains in the system being part of aggregates. Thus, for strongly micellizing systems such as T$_2$H$_8$T$_2$, there is a sizable temperature range over which aggregation is observed, but no first-order transition.  This is significant for interpreting the behavior of systems for which only aggregation is seen, discussed in the next subsection. 

Triblock chains with solvophobic outside blocks such as the two example architectures discussed here can have both of these attractive blocks be part of a single aggregate's solvophobic core, forming loops. Another option would be for the solvophobic blocks to bridge two different aggregates, which causes further aggregation. In experiments, such systems can lead to physical gelation as well as phase separation into chain-rich and chain-lean phases.\cite{Veg01}  

Another example of a system showing both aggregation and fluid-phase separation is the multiblock system (H$_3$T$_3$)$_3$. This system had been previously identified\cite{Gin08} as aggregating. We have shown here that the related (H$_3$T$_3$)$_4$ system, with one extra repeat unit, is phase-separating with no aggregates in the dilute phase prior to condensation. Simulations at multiple system sizes for (H$_3$T$_3$)$_3$ reveal a behavior qualitatively similar to that of the triblock chains discussed in this subsection. Data for this system are available as explained in the Data Availability Statement.  There is a broad region of volume fractions (or chemical potentials) in which aggregates form in an unsaturated dilute phase prior to reaching the binodal limit. This system, however, suffers from slow equilibration because of the multiple connection possibilities of the aggregates and the relatively low temperature at which the phase transition is observed. 

\subsection{Systems exhibiting only Aggregation}
Systems with relatively short hydrophobic blocks have been previously determined\cite{Flo99}  to aggregate into finite-size clusters, rather than phase separate into a bulk liquid phase. An example of this is H$_2$T$_4$, which was identified as aggregating, whereas H$_2$T$_8$ is confirmed to be phase separating only, as discussed in subsection III.A. 

\begin{figure}
    \centering
    \includegraphics[width=3.5 in]{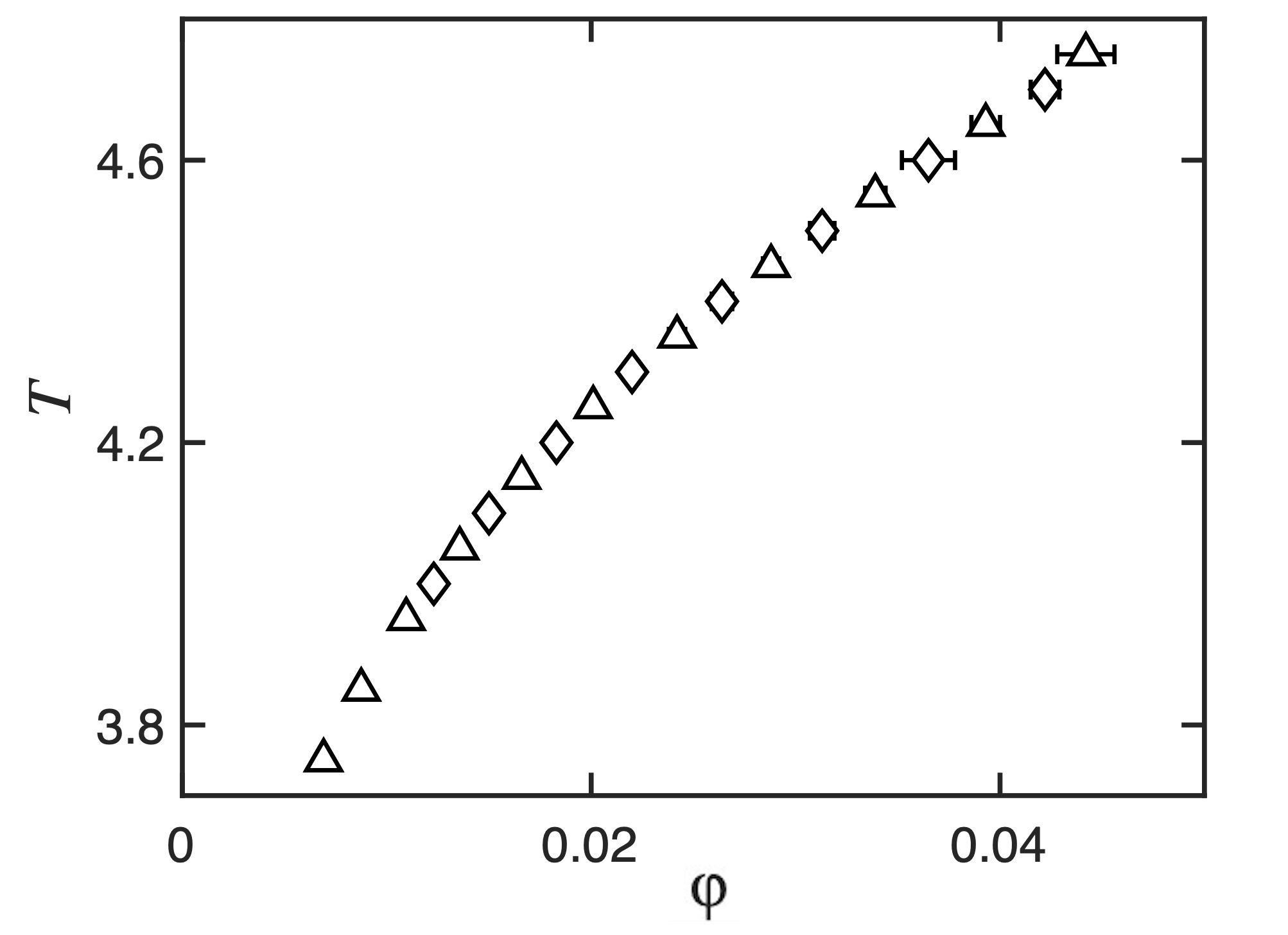}
    \caption{Critical micellar concentration for T$_2$H$_8$T$_2$ with $L=30$ (triangles) and $L=20$ (diamonds).}
    \label{fig:H2T4}
\end{figure}

The cmc curves obtained in the present work for H$_2$T$_4$ at two different system sizes are shown in Fig.~\ref{fig:H2T4}.  These cmc values agree well with those reported in Ref. \citenum{Flo99}. Because H$_2$T$_4$ has a relatively short hydrophilic segment that is not effective in protecting micelles from further aggregation, this system forms relatively large, disordered micellar aggregates at concentrations above the cmc. For example, at $T=4$, a run with $<\upvarphi>=0.015$, which is just 20\% above the cmc, in a box with $L=60$, showed a broad micellar peak with $<M>=85$. At higher concentrations, aggregates grow and become roughly cylindrical in shape. 

\begin{figure}
    \centering
    \includegraphics[width=3.5 in]{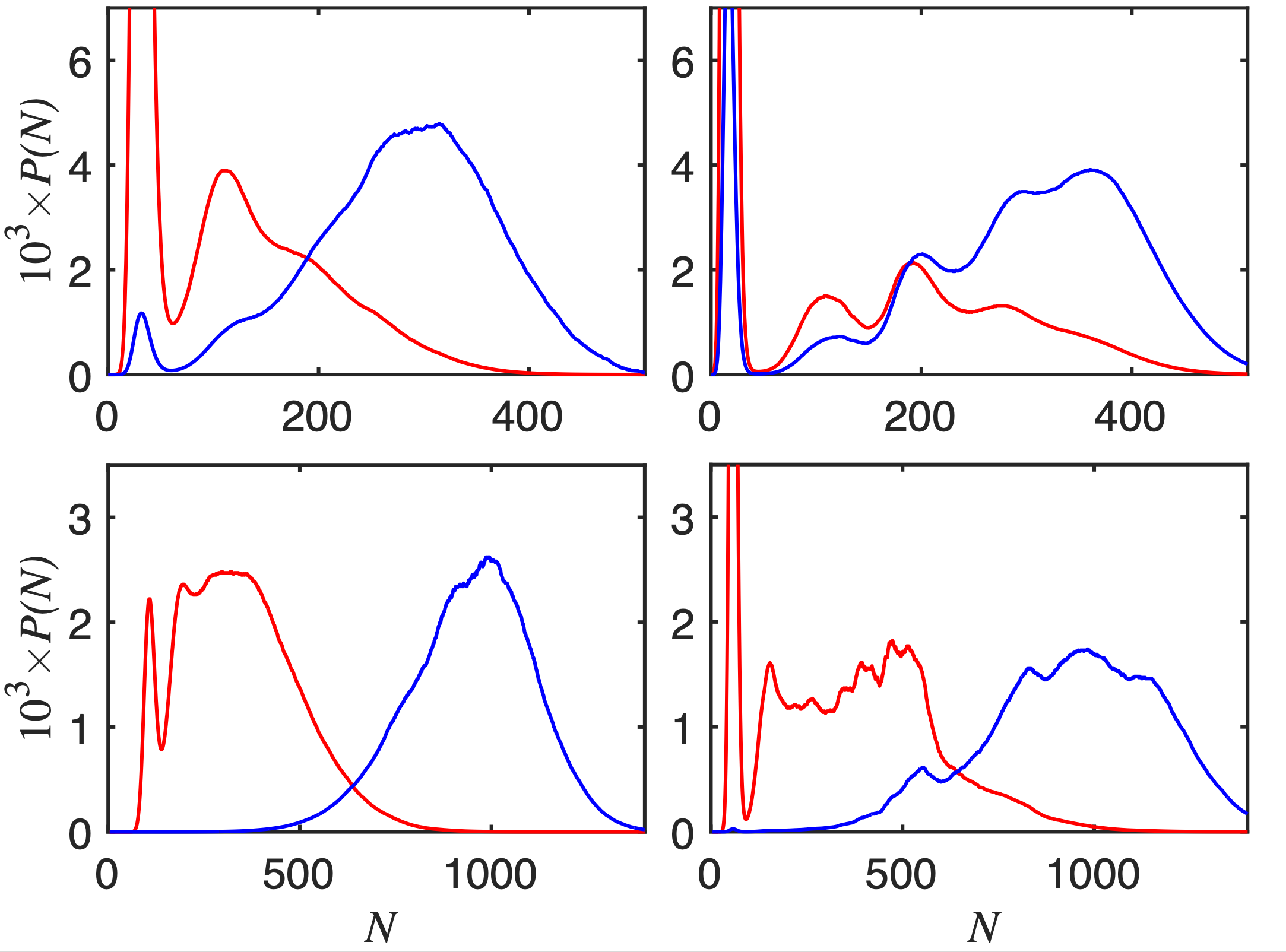}
    \caption{Number probability functions $P(N)$ for H$_2$T$_4$. Top row subplots are for $L=20$, bottom row for $L=30$. Left column subplots are at $T=4.3$, right column subplots are at $T=4.0$. The mean volume fractions are $<\upvarphi>=0.075$ for the red curves and $<\upvarphi>=0.21$ for the blue curves in all four subplots.}
    \label{fig:H2T4d}
\end{figure}

Probability distributions $P(N)$ versus $N$ for two temperatures ($T=4.3$ and 4) and two system sizes ($L=20$ and 30) are shown in Fig.~\ref{fig:H2T4d} for volume fractions $<\upvarphi>=0.075$ (red curves) and $<\upvarphi>$=0.21 (blue curves). These conditions are significantly above the cmc at either one of the two temperatures. The left-most peak, with is visible at the lowest concentration for both temperatures and system sizes, corresponds to oligomers-only present within the simulation box. The rest of the distribution corresponds to micellar aggregates of a very broad range of sizes. These number probability distributions are inconsistent in shape with Ising-type order parameter distributions (Fig.~\ref{fig:P_N} bottom), so there is no true critical point associated for H$_2$T$_4$ at any of the conditions studied. 

\begin{figure}
    \centering
    \includegraphics[width=3.5 in]{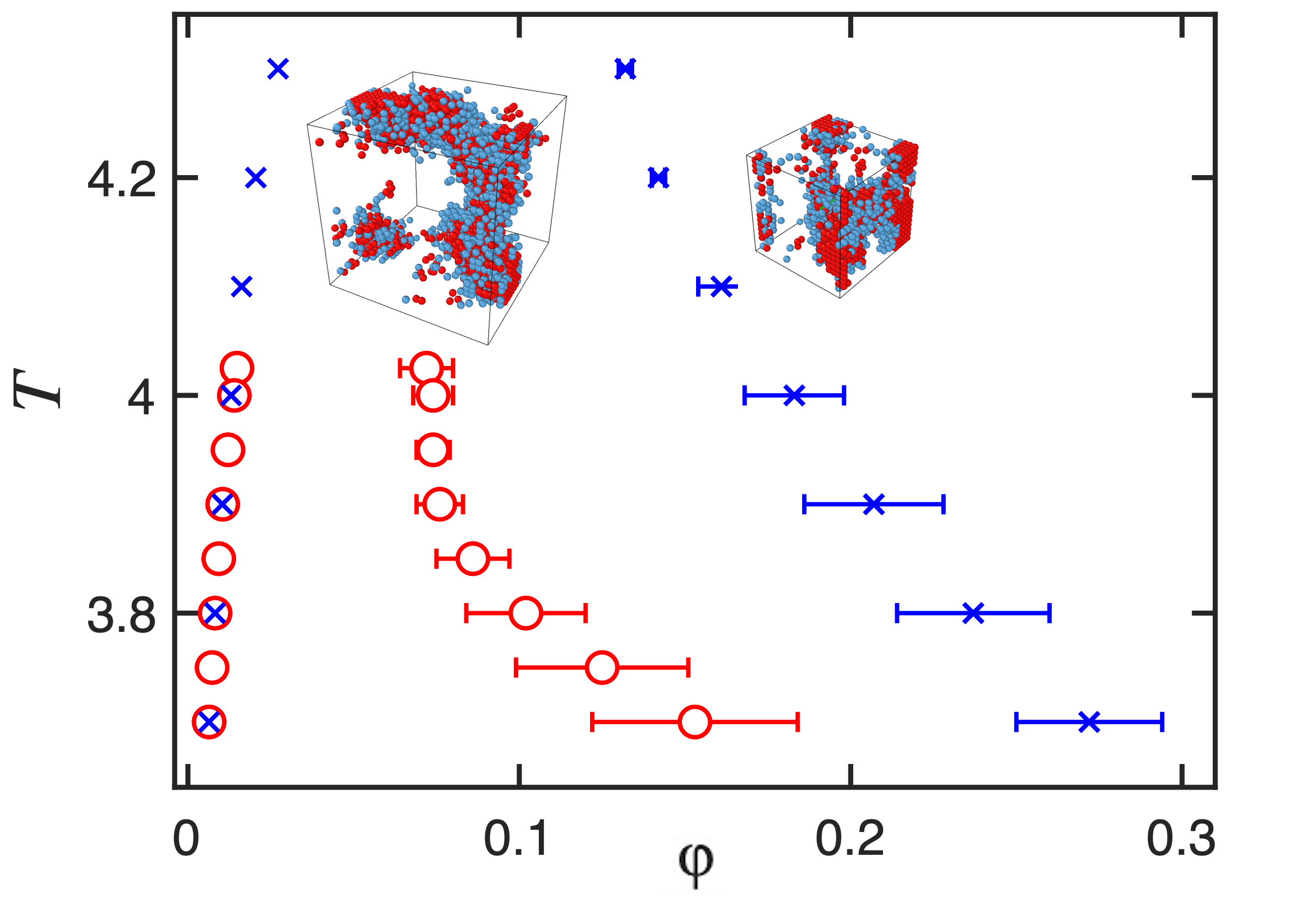}
    \caption{Pseudo-phase diagram for H$_2$T$_4$. The main plot shows volume fractions obtained from the area equality condition of the number probability functions for $L=30$ (red circles) and $L=20$ (blue $\times$'s). The inset shows representative configurations of the dense ``phase'' for the two system sizes.}
    \label{fig:H2T4p}
\end{figure}

The conclusion that there is no first-order phase transition for this system is also confirmed by forcing  computation of a ``pseudo-phase diagram'' through assigning the low-density peak to a putative dilute phase and the rest of the $P(N)$ distribution to a putative dense phase. The result of this calculation is shown in Fig.~\ref{fig:H2T4p}.  The apparent phase boundaries on the dilute-phase side are system-size invariant, but the dense-phase ``coexisting'' volume fractions are very strongly system-size dependent. The physical basis for this behavior can be seen from the representative configurations in the two different boxes shown within Fig.~\ref{fig:H2T4p}. The system-spanning irregular cylindrical micelles that form have roughly the same diameter, but occupy a bigger fraction of space in the smaller box. However, the chemical potential (volume fraction) at which the aggregates form is approximately independent of system size, so the dilute-phase boundary is insensitive to system size. The large statistical uncertainties seen for the dense-phase volume fractions are a consequence of slow dynamics for this system, due to the presence of system-spanning aggregates. 

The H$_2$T$_4$ system is representative of many chain architectures assigned previously to aggregation behavior. From continuity arguments with respect to systems such as T$_2$H$_8$T$_2$, for which we confirmed both aggregation and phase separation, we stipulate that the most likely fate of the first-order fluid-phase transition and associated Ising-type critical point is that the transition moves to lower temperatures than the ones at which micellization is readily observed. Any bulk liquid will have to be generated from attractive aggregate-aggregate interactions mediated by chain segments that have a strong preference for the hydrophobic cores of the aggregates. Clearly, the dynamics of these systems slow down as temperature is reduced. As a result, we are unable to observe a true first-order transition for many aggregating chain architectures.


\section{Conclusions}
In this work, we have re-examined the previously proposed distinction between macroscopic phase separation and formation of finite-size aggregates in systems of patterned linear chains consisting of hydrophobic and hydrophilic segments. By employing a careful analysis using grand canonical Monte Carlo simulations of a simple lattice mode, we were able to identify for the first time examples of both phase separation and aggregation behavior occurring in the same system at different thermodynamic conditions. Triblock chains with attractive outside segments are particularly prone to exhibiting this dual behavior, in agreement with experiments.\cite{Veg01} Short multiblock chains with a repeating hydrophobic-hydrophilic pattern are also shown here to exhibit this behavior.

In chain systems with attractive interactions, phase separation into  macroscopic dilute and dense fluid phases is a universal feature, taking place for the majority of sequences for long chains.\cite{Ran21} The exception to this behavior is chains with significant blocks of solvophilic beads, or very short chains, for which finite-size aggregates can be sterically prevented from attracting other aggregates and forming a macroscopic liquid, as confirmed in the present work. Aggregation behavior, which takes place at relatively low concentrations, was shown here to be possible at temperatures higher than the critical point for the fluid first-order phase transition. 

We hypothesize based on our results that at sufficiently low temperatures, all systems, even those found to exhibit ``aggregation only'' behavior, have a first-order phase transition between a dilute fluid phase and a dense liquid. The micellization line, which marks formation of a separate aggregate peak in the probability function $P(M)$ versus size $M$, and the binodal for the dilute phase, which marks the formation of the infinite-extent liquid, may be quite far apart for some chain architectures. Physically, this is due to the weakness of attractive interactions between aggregates because of the presence of solvophilic segments on their surface. However, it is expected that dynamic slow-down could prevent the actual observation of a first-order transition at temperatures for which a given system can reach equilibrium, in computer simulations or in the laboratory. Multiblock chains were found to be particularly prone to such non-equilibrium structure formation.   

\section*{Acknowledgements}

Financial support for this work was provided by the Princeton Center for Complex Materials (PCCM), a U.S. National Science Foundation Materials Research Science and Engineering Center (Award DMR-2011750). 

\section*{Data Availability Statement}
Computer codes used in this work, example input and output files, information on the runs performed, and numerical data for the phase boundaries and cmc's are freely available for download from the Princeton Data-Space repository, at DOI 10.34770/ykvp-8b36. 

\section*{References}
\bibliography{library}

\end{document}